\documentclass[12pt,a4 paper, one side,varwidth]{article}
\usepackage{amsmath,amssymb,latexsym}
\usepackage{hyperref}
\usepackage{graphicx}
\usepackage{subcaption}
\usepackage{epstopdf}

\begin{document}
\date{}
\title{Dynamics of expansion free self gravitating cylindrically symmetric radiating star  }
\author{Rajesh Kumar\footnote{rkmath09@gmail.com}\footnote{rajeshkumar.mathstat@ddugu.ac.in} and Sudhir Kumar Srivastava\footnote{sudhirpr66@rediffmail.com}\\
Department of Mathematics $\&$ Statistics,\\
Deen Dayal Upadhyaya Gorakhpur University, Gorakhpur, INDIA.}
\maketitle
\begin{abstract}
The present work  deals with the dynamics of radiating star which is considered to be expansion free  cylindrical symmetric dust dissipative fluids. Several treatments are adopted for the  description of geometrical and physical features of such stars. Firstly, it is shown that, the dynamical star does not permit the streaming out limit and diffusion approximation and also proved that acceleration and dissipation are necessary for  its dynamical evolution. It has also shown that, a static expansion free cylinder must be non-radiating. Secondly, the existence of cavity model and self-similar solution for the such dynamical star are also investigated.
\end{abstract}
Mathematics Subject Classification 2010: 83C05, 83F05, 83C75. \\
PACS numbers: 04.40.-b, 04.20.-q, 04.40.Dg, 04.40.Nr\\
Keywords: Expansion-free; Cylindrically symmetric; Dissipative fluids, Cavity evolution.
\section{Introduction}
The explosions in self-gravitating fluid distribution (stars) are significant phenomenon in relativistic astrophysics~\cite{mb09} . Over the years, there has been an extensive studies that how the collapsing system evolves after the explosion. An explosion in the center leads to an overall expansion of the fluid thus making a cavity surrounding the center. The pioneer work of Skripkin~\cite{s60}  revealed the fascinating phenomenon of cavity formation (vanishing expansion scalar) by assuming non-dissipation, constant energy density and isotropic pressure. The self gravitating relativistic system possess through various phase of evolution concerning kinematical quantities which determine the scenario of general relativistic gravitational collapse of massive star. The expansion scalar is one of these quantities which usually determine the rate of change of elementary volume fluid element.  

\par
In 2008, Herrera et al. \cite{lna08} studied the physical meaning of expansion-free motion and prove that this condition necessarily entail the appearance of a vacuum cavity (Minkowskian space-time) surrounding the center of fluid distribution. They have shown that in the process of contraction (expansion), the decrease (increase) in a volume due to the decreasing (increasing) area of the external boundary surface must be balanced  by decrease(increase) of the internal boundary surface (cavity creation) under expansion-free motion.\\
Some analytical solutions for the evolution of expansion free non-dissipative anisotropic fluids has been  discussed by Di Prisco et al. \cite{dhosv11}.  Some recent works illustrated the dynamical stability of the self gravitating expansion-free fluid distribution ( \cite{dhosv11} - \cite{sa14} and references their in ). Sharif and Yousaf \cite{sy12} have discussed the cylindrically symmetric non-dissipative inhomogeneous expansion free model and showed that the solution is completely integrable. Sharif and Yousaf (\cite{sy12} - \cite{sy12a}) devised the expansion-free analytical models for cylindrically symmetric and plane symmetric case in non-dissipative fluid. Herrera et al. \cite{hds09} investigated that the analytical solutions for cavity evolution with inhomogeneous energy density distributions, which are relatively simple to analyze but still may contain some of the essential features of a realistic situation. 

\par
Thus the evolution of expansion-free self gravitating system offers interesting astrophysical consequences and this scenario allows for the obtention of a wide range of models for the self gravitating fluid distribution.
In this paper, authors are interested to investigate the dynamics of radiating cylindrically symmetric star and discuss the analytical solutions of field equations.

\section{Basic Equations: the interior metric, cavity evolution and Einstein's field equations}
We consider a collapsing radiating cylinder filled with dust undergoing dissipation in the form of heat flow and  free-streaming radiation bounded by hypersurface $\Sigma^e$, which divides the space-time into two distinct 4-dimensional manifolds $V^+$ and $V^-$. 
\subsection{The interior metric and cavity evolution}
For the bounded configuration, a line element applies to the 
fluid inside $\Sigma^e$, choosing a comoving coordinate the general cylindrically symmetric metric can be written as
\begin{equation}
ds_-^2 = - A^2 dt^2 + B^2 dr^2 + C^2(d\phi^2 + dz^2)
\label{eq1}
\end{equation}
where the metric coefficient $A, B$ and $C$ are function of (t,r) and coordinate labels $x_-^i = (t,r, \phi, z), i =0,1,2,3$. In order to represent cylindrical symmetry, the range of coordinates is required to be as follows:
$$ -\infty < t < +\infty, ~\quad~ 0 \leq r < + \infty, $$
$$ 0 \leq \phi  \leq 2 \pi, ~\quad~ -\infty < z < +\infty $$
The  fluid distribution is assumed to be  dust undergoing dissipation, represented by the energy momentum tensor (inside $\Sigma^e$ if the system is bounded) to have the form
\begin{equation}
T^i_j = \rho v^i v_j + v^i q_j  + q^i v_j + \epsilon \ell^i \ell_j
\label{eq2} 
\end{equation}
 where, $\rho\rightarrow$ energy density, $q^i \rightarrow$  heat flux, $\epsilon \rightarrow$ radiation density, $v^i \rightarrow$ four velocity vector of the fluid and $\ell^i \rightarrow$ null four vector.\\
 
 For the comoving coordinate system these satisfying
\begin{equation}
v^i = A^{-1} \delta^i_0, \quad~ \ell^i = A^{-1} \delta^i_0  + B^{-1} \delta^i_1, ~\quad~ q^i = q B^{-1} \delta^i_1 
\label{eq3}
\end{equation} 
such that
\begin{equation}
v^i v_i  = -1,~ \quad~ \ell^i v_i = -1,~\quad~ \ell^ \ell_i = 0, ~\quad~ q^i v_i = 0
\label{eq4}
\end{equation}
The shear tensor $\sigma_{ij}$, acceleration vector $a_i$ and the expansion scalar $\theta$ are defined as
\begin{equation}
\sigma_{ij} = \frac{1}{2} (v_{i;j} + v_{j;i})+ \frac{1}{2} (a_i v_j + a_j v_i) -\frac{1}{3} \theta (g_{ij} + v_i v_j)
\label{eq5}
\end{equation}
\begin{equation}
a_i = v_{i;j} v^j ~\quad~ \mbox{and} ~\quad~ \Theta = v^i_{;i}
\label{eq6}
\end{equation}
Now, it follows from equs.(\ref{eq3}) - (\ref{eq6}) that
\begin{equation}
\sigma_{11} = \frac{2}{\sqrt{3}} B^2 \sigma, ~\quad~ \sigma_{22} = -\frac{1}{\sqrt{3}} C^2 \sigma = \sigma_{33}
\label{eq7}
\end{equation}
where
\begin{equation}
\sigma = \frac{1}{\sqrt{3}}\frac{1}{A} (\frac{\dot{B}}{B} - \frac{\dot{C}}{C})
\label{eq8}
\end{equation}
where $\sigma^2 = \frac{1}{2} \sigma_{ij} \sigma^{ij}$. 
\begin{equation}
a_i = \frac{A'}{A} \delta^1_i
\label{eq9}
\end{equation}
\begin{equation}
\Theta = \frac{1}{A} (\frac{\dot{B}}{B} + 2 \frac{\dot{C}}{C})
\label{eq10}
\end{equation}
Here, dash ($'$) and over-dot ($.$) refer to derivative with respect to $r$ and $t$ respectively.
\par
A cylindrically symmetric space-time defined locally by the existence of two commuting, space-like Killing vectors $( \xi_\phi, \xi_z) = (\frac{\partial}{\partial\phi}, \frac{\partial}{\partial z})$  and  the existence of cylindrical symmetry about an axis entails that the orbits of one of these vectors are closed but those of the other is open~\cite{c96}. Each of these killing vectors must be hypersurface orthogonal. The norms of these Killing vectors are invariants, namely the circumferential radius  $\kappa$  and specific length $\eta$ ~(see ~\cite{c96} - \cite{h00} for more details)
\begin{equation*}
\kappa = \sqrt{\xi_\phi \xi^*_\phi}  = \sqrt{\xi_{(2)i} \xi_{(2)}^i}
\end{equation*}
\begin{equation*}
\eta = \sqrt{\xi_z \xi^*_z}  = \sqrt{\xi_{(3)i} \xi_{(3)}^i}
\end{equation*}
where $\xi_{(2)}= \partial_\phi, \xi_{(3)}= \partial_z$ and $\xi^*$ represents the dual vector with respect to metric (\ref{eq1}). The gravitational energy per specific length in a cylindrically symmetric system (also known as $C-energy$) is defined by~\cite{t65}
\begin{equation}
\mathcal{E} = \frac{1}{8} (1- \eta^{-2} \nabla^i \hslash \nabla_i \hslash), ~\quad~ \hslash = \kappa \eta
\label{eq11c}
\end{equation}
If the evolution of cylindrically symmetric fluid is assumed to be expansion-free i.e., $\Theta = 0$. Then it follows from equ.(\ref{eq10}) that
\begin{equation}
B = \frac{1}{C^2}
\label{eq11a}
\end{equation}
Then equ.(\ref{eq1}) reduces to the space-time metric for cavity evolution\cite{lna08}
\begin{equation}
ds_-^2 = - A^2 dt^2 + C^{-4} dr^2 + C^2(d\phi^2 + dz^2)
\label{eq12}
\end{equation}
In this paper author is interested in the dynamics of self gravitating system  only in its cavity evolution once it is already formed.
\par

The Einstein's field equation
\begin{equation*}
G^i_j = \frac{8 \pi G}{c^4} T^i_j
\end{equation*}
for the system (\ref{eq12}) yields
\begin{equation}
-8\pi \tilde{\rho} = 5 C'^2 C^2 + 2 C'' C^3 + \frac{3}{A^2}\frac{\dot{C}^2}{C^2}
\label{eq13}
\end{equation}
\begin{equation}
4 \pi \tilde{q} A = 2 C' \dot{C} + C \dot{C'} -C \dot{C} \frac{A'}{A}
\label{eq14}
\end{equation}
\begin{equation}
8 \pi \epsilon C A^2 = A^2 C^3 C'^2 + 2 A C^4 A' C' + 2\frac{\dot{A}}{A} \dot{C} -\frac{\dot{C}^2}{C}  -2 \ddot{C}
\label{eq15}
\end{equation}
\begin{equation}
\ddot{C} + A^2 C^3 (2 C'^2 + C C'') + A^2 C^5 (\frac{3 A' C'}{A C} + \frac{A''}{A})-\dot{C} (\frac{\dot{A}}{A}+4\frac{\dot{C}}{C})=0
\label{eq16}
\end{equation}
where $\tilde{\rho} = \rho + \epsilon, \tilde{q} = q+\epsilon$ and taken $G \sim 1, c \sim 1$. The non-vanishing components of the Bianchi identities $T^{ij}_{;j} = 0$ are
\begin{equation}
\frac{\dot{\tilde{\rho}}}{A} - \frac{2\epsilon}{A} \frac{\dot{C}}{C} + 2 \tilde{q} C^2 (\frac{A'}{A} + \frac{C'}{C}) + C^2 \tilde{q}'==0
\label{eq17}
\end{equation}
\begin{equation}
\frac{\dot{\tilde{q}}}{A} - \frac{2 \tilde{q}}{A} \frac{\dot{C}}{C} + (C^2 \epsilon)' + C^2 (\tilde{\rho}+\epsilon)\frac{A'}{A}=0
\label{eq18}
\end{equation}
Also, in view of (\ref{eq8}) and (\ref{eq11a}), equ. (\ref{eq14}) yields
\begin{equation}
4 \pi \tilde{q} + \frac{1}{\sqrt{3}} \sigma' C^2 + \sqrt{3} \sigma C C' = 0
 \label{eq19}
\end{equation}
The fluid velocity is given by $U = \frac{\dot{C}}{A}$ which must be negative to ensure the collapsing dust cylinder. The gravitational energy defined by (\ref{eq11c}) then becomes
\begin{equation}
\mathcal{E} = \frac{C}{2} (\frac{\dot{C}^2}{A^2} - C^4 C'^2)+ \frac{1}{8}
\label{eq20}
\end{equation}
The t-rate of variation of the total energy inside the  cylinder is
\begin{equation}
\dot{\mathcal{E}} = 4\pi C^2 A [ -U(\epsilon-\frac{1}{32 \pi C^2}) - C' C^2 \tilde{q} ]
\label{eq21}
\end{equation}
It follows from (\ref{eq21}) that in the collapsing process ($U<0$) the coefficient of $U$ will increase the $C-energy$ of the cylinder if $\epsilon > \frac{1}{32 \pi C^2} $ (i.e., the radiation energy is greater than a fixed value). Hence,the presence of  dissipation energy leads to the increase of $C-energy$. The second term in square bracket, due to the negative sign, describe the out flow of $C-energy$ in the form of heat flux.

\par 
The r-variation of energy between the adjacent co-axial cylinder inside the fluid is
\begin{equation}
\mathcal{E'} = 4 \pi C' C^2 (\tilde{\rho} + \tilde{q} C'C^2 U)
\label{eq22}
\end{equation}
The first term on right hand side provides the contribution of energy density of fluid element inside a cylindrical shell along the heat flux and radiation. However, since $U<0$ then second term decreases the energy of cylinder during the process.

\par
The Weyl Scalar $W$ in context of  Kretchman scalar $\Upsilon$ is defined as
\begin{equation}
W^2 = \Upsilon - 2 R^{ij} R_{ij} + \frac{1}{3} R^2
\end{equation}
where $R$ is scalar curvature and the Kretchman scalar $\Upsilon = R^{ijkm} R_{ijkm}$ yields
\begin{equation}
\Upsilon = \frac{16}{C^6} (\mathcal{E} - \frac{1}{8})[ 3(\mathcal{E} - \frac{1}{8})- 8 \pi \rho C^3 ] + (8 \pi)^2 [ 3\rho^2 -4 q^2 + 4\epsilon(\rho-2q) ]
\label{w1}
\end{equation}
In view of (\ref{w1}), the Weyl scalar $W$  for the metric (\ref{eq12}) takes the form
\begin{equation}
W^2 = \frac{48}{C^6} (\mathcal{E} - \frac{1}{8})^2+ 8 \pi \rho [ \frac{56}{3} \pi \rho - \frac{16}{C^3} (\mathcal{E} - \frac{1}{8}) ]
\label{w2}
\end{equation}
which is the relation between the Weyl tensor and energy density inhomogeneity of the fluid and it can be seen  that dissipation of expansion free cylinder do not affect the Weyl-geometry of space time.

\subsection{The exterior metric and junction conditions} 
The collapsing fluid reside inside the metric (\ref{eq1})  must be matched to a suitable exterior space-time. If the radiation (dissipation)  leaves the fluid across boundary surface then exterior region $V^+$ of the collapsing star will not be vacuum, but the outgoing \textit{Vaidya like space-time} which model the radiating star~\cite{cg95}. Then, for the junction conditions, consider  the Vaidya's metric (\cite{cg95}, \cite{v53}) for the exterior region $V^+$
\begin{equation}
ds^2_+ = -(\frac{-2M(\tau)}{\Re})-2 d\Re d\tau + \Re^2 (d\phi^2+ dz^2)
\label{j1}
\end{equation}
where coordinates are label as  $x^i_+ = (\tau,\Re, \phi, z)$. $M$ and $\tau$ described are the total mass and retarded time respectively.

\par

For smooth matching of the interior ($\Sigma^-$) and exterior($\Sigma^e$) regions, Darmois condition lead to (see \cite{anc85}, \cite{sa17} for more details)
\begin{equation}
\mathcal{E} =^{\Sigma^e}  M(\tau) + \frac{1}{8}, ~\quad~ q =^{\Sigma^e} 0
\label{j2}
\end{equation}
$=^{\Sigma^e}$ indicates that the quantities are evaluated at external hypersurface. Since the expansion-free case delimits a vacuum cavity (Minkowski space-time)  which ensure a hypersurface $\Sigma^i$ separating  fluid distribution and the central Minkowskian cavity from the collapsing fluid \cite{dhosv11}. Then the junction condition over hypersurface $\Sigma^i$,
\begin{equation}
\mathcal{E} =^{\Sigma^i} 0, ~\quad~ q =^{\Sigma^i}  0
\label{j3}
\end{equation}
The total luminosity of the collapsing matter visible to an observer at rest at infinity is \cite{anc85}
\begin{equation*}
L_\infty = -(\frac{dM}{d\tau})_{\Sigma^e}
\end{equation*}
which gives
\begin{equation}
L_\infty = 4 \pi \epsilon C^2 (\frac{\dot{C}}{A}+ C' C^2)^2
\label{j4}
\end{equation}
Thus, the total luminosity of the fluid as visible to the distant observer depends on the radiation energy only. For an observer at the boundary surface $\Sigma^e$, the luminosity is (\cite{anc85}-\cite{ gb14})
\begin{equation}
L_{\Sigma^e}= -[(\frac{d\tau}{d\varsigma})^2 \frac{dM}{d\tau}] =4 \pi \epsilon C^2
\label{j5}
\end{equation}
$\varsigma$ is intrinsic time coordinate over $\Sigma^e$.

\par
The boundary red-shift of the radiation emitted by the collapsing fluid can be measured as
\begin{equation*}
Z_{\Sigma^e} = \sqrt{\frac{L_{\Sigma^e}}{L_\infty} -1} 
\end{equation*}
which yields
\begin{equation}
1+Z_{\Sigma^e} = (\frac{\dot{C}}{A} +C' C^2)^{-1}
\label{j6}
\end{equation}
It follows from (\ref{j4} - \ref{j6}) that 
\begin{equation}
L_{\infty} = \frac{L_{\Sigma^e}}{(1+Z_{\Sigma^e})^2}
\label{j7}
\end{equation}
Thus, the luminosity measured by an observer rest at infinity is reduced by the red- shift in comparison to the luminosity observed at the surface $\Sigma^e$.


\section{Dissipative approximation: Streaming out limit and Diffusion approximation}
The gravitational collapse of star is a highly dissipative process and  is usually treated appealing into two approximations namely  diffusion-approximation ($\epsilon=0, q\neq 0$ ) and streaming-out limit( $q=0, \epsilon \neq 0$ ) \cite{is76}-\cite{l88}. The former one is in general very sensible, since it applies whenever the mean free path of particles responsible for the propagation of energy is very small as compared with the typical length of the object, a fact found very often in astrophysical scenarios while when the mean free path of particles transporting energy may be large enough so as to justify the streaming-out approximation. Thus, it consist of concurrently both limiting cases of radiative transport (diffusion and streaming out), allowing one to illustrate a wide range of situations.

\par
If the fluid distribution is in the streaming out limit approximation($q =0, \epsilon \neq 0$), it follows from equs (\ref{eq17})-(\ref{eq18})  that
\begin{equation}
\rho = \rho_0(r) e^{\int A' C^2 dt}
\label{d3}
\end{equation}
$\rho_0$ is an arbitrary. If the motion of cylindrical fluid is of zero acceleration then equ. (\ref{d3}) reveals that $\rho = \rho(r)$, a static configuration of the system.
\par

Further, for  the diffusion approximation ($q\neq 0, \epsilon = 0$), it can been seen from equ. (\ref{eq19}) that
\begin{equation}
4 \pi q = -\frac{1}{\sqrt{3}}  (\sigma' C^2 + 3 \sigma C C')
\label{d4}
\end{equation}
The equ.(\ref{d4}) reveals that heat flux of expansion free dynamical star is occurs due to the shear of the fluid. Also, $\sigma = 0 \Rightarrow 4 \pi q = 0$ which showed that static expansion free cylinder must be non-radiating.

\par
However, in particular for non-dissipative case ( \cite{rs18}-\cite{rs18a} ), we have from (\ref{d4})
\begin{equation}
\sigma = \frac{k(t)}{C^3}
\label{d5}
\end{equation}
where $k(t)$ is arbitrary function.\\
If we consider the motion of fluid distribution is geodesic ($A' = 0$), it follows from Equs. (\ref{eq17})-(\ref{eq18}) that 
\begin{equation}
q = q_0(r) C^2
\label{d6}
\end{equation}
\begin{equation}
 8 \pi \dot{\rho} + (8 \pi q C^2)' = 0
 \label{d7}
\end{equation}
$q_0$ is an arbitrary. By, applying the boundary condition (\ref{j3})  we obtain $ q_0 (r) = 0 $ and then we have $q = 0$ and from (\ref{d7}), $\rho = \rho(r)$. Therefore it must be $A' \neq 0$ and $q \neq 0$ for dynamical evolution of  expansion-free cylindrical star.
\par
From the above results it is clear  that an expansion  free dynamical cylinder(non-static) does not allows streaming out limit and diffusion approximation in geodesic case. 

\par
Thus, it also can be stated that- self gravitating expansion free dynamical cylinder must be accelerating and dissipating simultaneously. Similar results has also been obtained by Sherif et al. \cite{sgm19} for LRS-II class space time.

\section{Model of dynamical star}
Motivated by the results from  previous section, the authors are interested to discuss the dynamical (accelerating and dissipating) model of cavity evolution.  Considering the ansatz referred in \cite{pc13} consisting subsequent form of metric coefficient as separate function of coordinate $t$ and $r$ as,
\begin{equation}
A(t,r) = \mathcal{A} (r), \quad~ C(t,r) = \mathcal{A} (r) \mathcal{C}(t)
\label{ng1}
\end{equation}
where $\mathcal{A} (r)$ describes static configuration  with energy density $\rho_0(r)$ given by
\begin{equation}
4 \pi \rho_0(r) = \mathcal{A}^4 ( \frac{\mathcal{A''}}{\mathcal{A}}+4\frac{\mathcal{A'}^2}{\mathcal{A}^2} )
\label{ng2}
\end{equation}
Thus in view of  (\ref{eq20}) and (\ref{ng1})
\begin{equation}
\mathcal{E} = \frac{1}{8}+ \frac{\mathcal{A} \mathcal{C}}{2} ( \dot{\mathcal{C}}^2 - \mathcal{A'}^2 \mathcal{A}^2 \mathcal{C}^6 )
\label{ng6}
\end{equation}
It follows from equ. (\ref{eq16}) that
\begin{equation}
\frac{1}{\mathcal{C}^4} (\frac{\ddot{\mathcal{C}}}{\mathcal{C}} - 4\frac{\dot{\mathcal{C}}^2}{\mathcal{C}^2}) + (5\mathcal{A'}^2 \mathcal{A}^4 + 2\mathcal{A''} \mathcal{A}^5) = 0
\label{ng7}
\end{equation}
which gives after integration
\begin{equation}
\mathcal{C}(t) = \frac{k_2}{(k_1+3t)^{\frac{1}{3}}}, ~\quad~ \mathcal{A}(r) = k_4(7r-2k_3)^{\frac{2}{7}}
\label{ng9}
\end{equation}
where $k_1, k_2, k_3$ and $k_4$ are integrating functions. Therefore, the line element (\ref{eq12}) takes the form
\begin{equation}
ds^2 =  (7r-2 k_3)^{\frac{4}{7}} [ -dt^2 + (k_1+3t)^{\frac{4}{3}} (7r-2k_3)^{-\frac{12}{7}} dr^2+ (k_1+3t)^{-\frac{2}{3}} (d\phi^2 + dz^2) ]
\label{nga9}
\end{equation}
where the  constants are taken as $k_2 \rightarrow 1, k_4 \rightarrow 1$.
\par
Thus by virtue of (\ref{ng9}), equ.(\ref{eq13})-(\ref{eq15}) yield
\begin{equation}
8 \pi \rho = -8 \pi \rho_0 + \frac{12}{(k_1+3t)^2 (7r-2k_3)^{\frac{4}{7}}}
\label{ng9a}
\end{equation}
\begin{equation}
8\pi q = \frac{1}{(7r-2k_3)^{\frac{6}{7}} (k_1+3t)^2} [ 9(7r-2k_3)^{\frac{2}{7}} - 4 (7r-2k_3)^{\frac{1}{7}}(k_1+3t)^{\frac{1}{3}}-12(k_1+3t)^{\frac{2}{3}} ]
\label{ng9b}
\end{equation}
\begin{equation}
8 \pi \epsilon = \frac{3}{(7r-2k_3)^{\frac{6}{7}} (k_1+3t)^2} [ 4 (k_1+3t)^{\frac{2}{3}} - 3 (7r-2k_3)^{\frac{2}{7}} ]
\label{ng9c}
\end{equation}

Also, from Equs. (\ref{eq8}) and (\ref{eq9})
\begin{equation}
\sigma = \sqrt{3} (k_1 + 3t)^{-1} (7r-2k_3)^{\frac{-2}{7}}
\label{ng9d}
\end{equation}
\begin{equation}
a_i = (\frac{2}{7r-2k_3}) \delta^1_i
\label{ng9e}
\end{equation}
 The gravitational energy and boundary red-shift of radiation are 
\begin{equation}
\mathcal{E} - \frac{1}{8} = \frac{1}{2(k_1+3t)^3} [ (7r-2k_3)^{\frac{2}{7}} - 4 (k_1+3t)^{\frac{2}{3}} ]
\label{ng10}
\end{equation}
\begin{equation}
1+Z_{\Sigma^e} = \frac{(k_1+3t)^{\frac{4}{3}} (7r-2k_3)^{\frac{1}{7}}}{2 (k_1+3t)^{\frac{1}{3}} - (7r-2k_3)^{\frac{1}{7}}}
\label{ng11}
\end{equation}

\section{Self-Similar Regime}
 The ideas of self-similarity has a significant role in the extensively studies of Newtonian and relativistic problems. The self similar solutions of field equation in general relativity were widely studied in details (for more detail see \cite{ct71} - \cite{w06}).  The existence of a similarity (self-similar) solution of the field equation as one for which the resulting space-time admits the homothetic Killing vectors $\xi^i$ satisfying
 \begin{equation}
 \mathcal{L}_\xi g_{ij} = 2g_{ij}
 \label{ss1}
 \end{equation}
 The advantage of considering self-similar regime is that the field equation, a set of partial differential equation reduces into ordinary differential equation. In this section, it has studied the appearance of a homothetic Killing vector field for cylindrically symmetric space-time entails the detachability of the metric components in terms of comoving coordinate and that the line element can be expressed in a simplified unique form.
 
 \par
Consider a homothetic vector $\xi^i$ $(i= 0, 1, 2, 3)$ which commute with the two Killing vectors i.e., the Abelian similarity group $H_3$ there are other homothetic vector \cite{cms94}, but it is focused on the one as it is cylindrically symmetric

\begin{equation}
\xi^i  = \alpha(t, r) \delta^i_t + \beta(t, r) \delta^i_r
\label{ss2}
\end{equation}

If the line element (\ref{eq12}) admits homothetic Killing vector of the form (\ref{ss2}), the Equ. (\ref{ss1}) yield the following system of equations
\begin{equation}
\beta A' + \alpha \dot{A} + A (\alpha-1)=0
\label{ss3}
\end{equation}
\begin{equation}
\dot{\beta} = A^2 C^4 \alpha'
\label{ss4}
\end{equation}
\begin{equation}
\beta' = 1+ \frac{2}{C} (\beta C' + \alpha \dot{C})
\label{ss5}
\end{equation}
\begin{equation}
\beta C' + \alpha \dot{C} - C=0
\label{ss6}
\end{equation}
In view of Equs. (\ref{ss5}) and (\ref{ss6})
\begin{equation}
\beta (t,r) = k_1 (t) + 3 r
\label{ss7}
\end{equation}
$k_1(t)$ is an arbitrary function. Using Equs. (\ref{ss3}) - (\ref{ss5}) in (\ref{ss6}) gives
\begin{equation*}
\alpha \frac{\dot{\alpha'}}{\alpha'} + (k_1 + 3r) \frac{\alpha''}{\alpha'} -2 \dot{\alpha}-\alpha \frac{\ddot{k_1}}{\dot{k_1}}+6=0
\end{equation*}
which has solution
\begin{equation}
\alpha(u) = -(3 c_2 + \frac{e^{-2c_1}}{u-c_2})
\label{ss8}
\end{equation}
where $u = t+r$ is defined by the new coordinate, $k_1(t) = 3t$ and $c_1, c_2$ are arbitrary constants. By virtue of Equs. (\ref{ss7}) - (\ref{ss8}), the equations (\ref{ss4}) and (\ref{ss6}) yield
\begin{equation}
A^2(u) = \frac{3 e^{2c_1} (u-c_2)^2}{c_3^4 (1-3e^{2c_1}(u-c_2)^2)^{\frac{2}{3}}}
\label{ss9} 
\end{equation}
\begin{equation}
C(u) = c_3 (1-3 e^{2c_1} (u-c_2)^2)^{\frac{1}{6}}
\label{ss10}
\end{equation}
where $c_3$ is arbitrary constant. Therefore, the line element (\ref{eq12}) reduces to the form
\begin{equation}
ds^2  = \chi^{\frac{-2}{3}} [ -(1-\chi) dt^2 + dr^2+ \chi(d\phi^2 + dz^2) ]
\label{ss11}
\end{equation}
where $\chi = 1-3(u-c_2)^2$ denotes the function of similarity variable $u = t+r$ and arbitrary constants are taken as $c_1 \rightarrow 0, c_3 \rightarrow 1$. 

\par
The physical quantities $\epsilon, q$ and $\rho$ can be obtained from the field equations (\ref{eq13}) - (\ref{eq15})
\begin{equation}
8 \pi \epsilon = \frac{1}{3} \chi^{\frac{-4}{3}} (2-3 \chi)
\label{ss12}
\end{equation}
\begin{equation}
4 \pi q = \frac{1}{3} \chi^{\frac{-4}{3}} (3\chi-2-2\sqrt{1-\chi})
\label{ss13} 
\end{equation}
\begin{equation}
8 \pi \rho = \frac{4}{3} \chi^{\frac{-1}{3}}
\label{ss14}
\end{equation}
From equs. (\ref{eq8}) and (\ref{eq8}) , the shear and acceleration are
\begin{equation}
\sigma = \chi^{\frac{-2}{3}}
\label{ss15}
\end{equation}
\begin{equation}
a_i = \frac{1}{\sqrt{3}} \frac{(\chi+2)}{\chi \sqrt{1-\chi}} \delta^1_i
\label{ss16}
\end{equation}
The gravitational energy from (\ref{eq20}),
\begin{equation}
\mathcal{E} = \frac{1}{6} \chi^{\frac{1}{6}}+\frac{1}{8}
\label{ss17}
\end{equation}
It also follows from equs. (\ref{j5}) and (\ref{j6}), the luminosity ($L_{\Sigma^e}$) and red-shift ($Z_{\Sigma^e}$) observed at the external surface 
\begin{equation}
L_{\Sigma^e} = \frac{1}{6} \frac{(2-3\chi)}{\chi}
\label{ss18}
\end{equation}
\begin{equation}
1+Z_{\Sigma^e} = -\frac{\sqrt{3\chi}}{1+\sqrt{1-\chi}} < 0 ~\qquad~ \mbox{(Blue-shift)}
\label{ss19}
\end{equation}
\begin{figure}[h]

\begin{subfigure}[b]{0.6\textwidth}
\includegraphics[width=\textwidth]{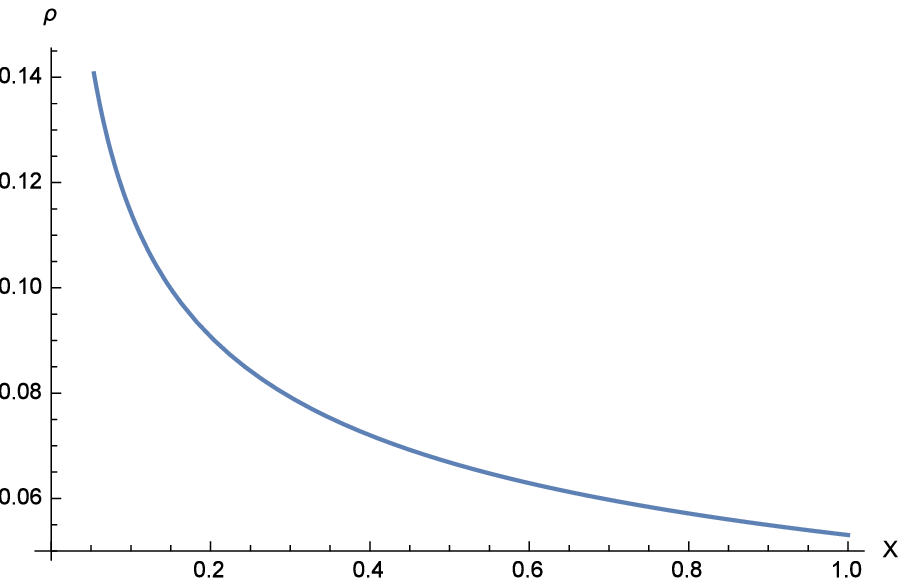}
        \caption{}
\end{subfigure}
\begin{subfigure}[b]{0.6\textwidth}
\includegraphics[width=\textwidth]{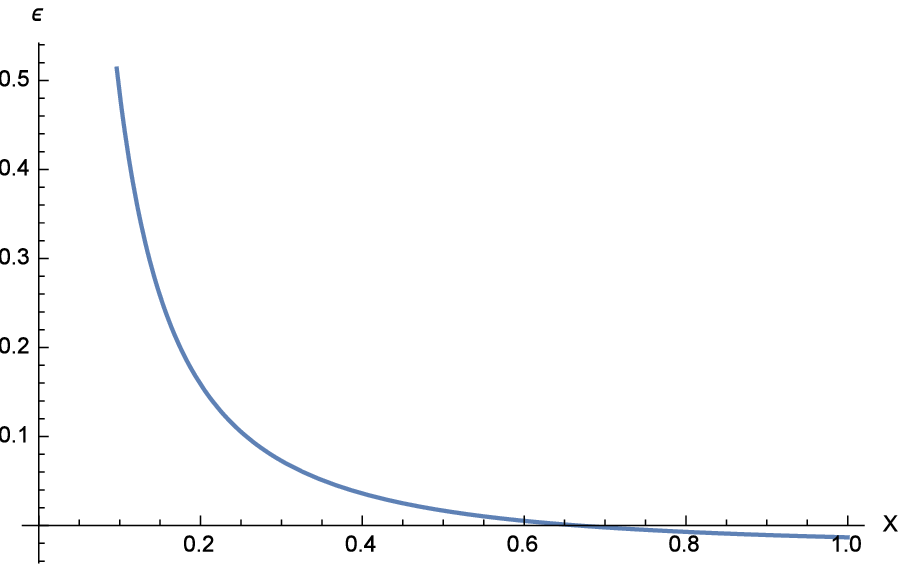}
        \caption{}
\end{subfigure}

\begin{subfigure}[b]{0.6\textwidth}
\includegraphics[width=\textwidth]{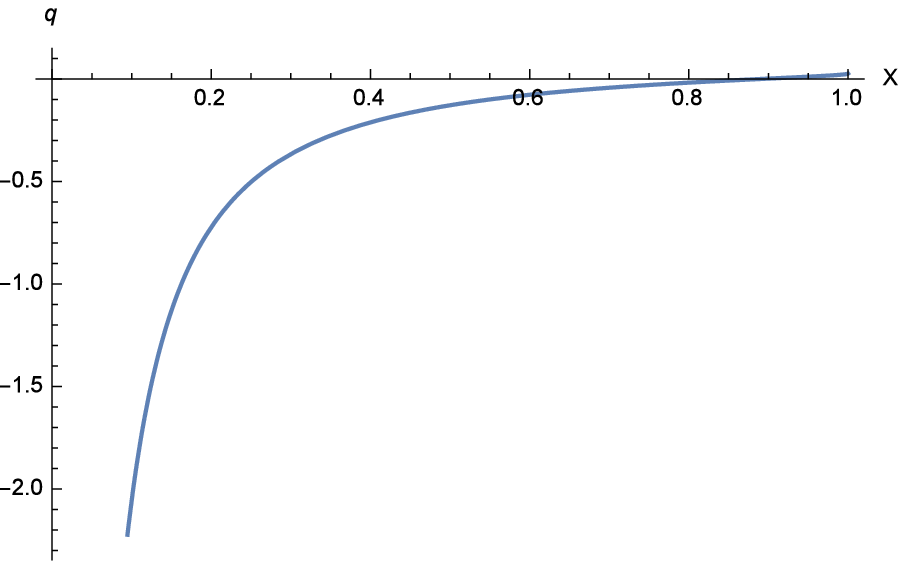}
        \caption{}
\end{subfigure}
\begin{subfigure}[b]{0.6\textwidth}
\includegraphics[width=\textwidth]{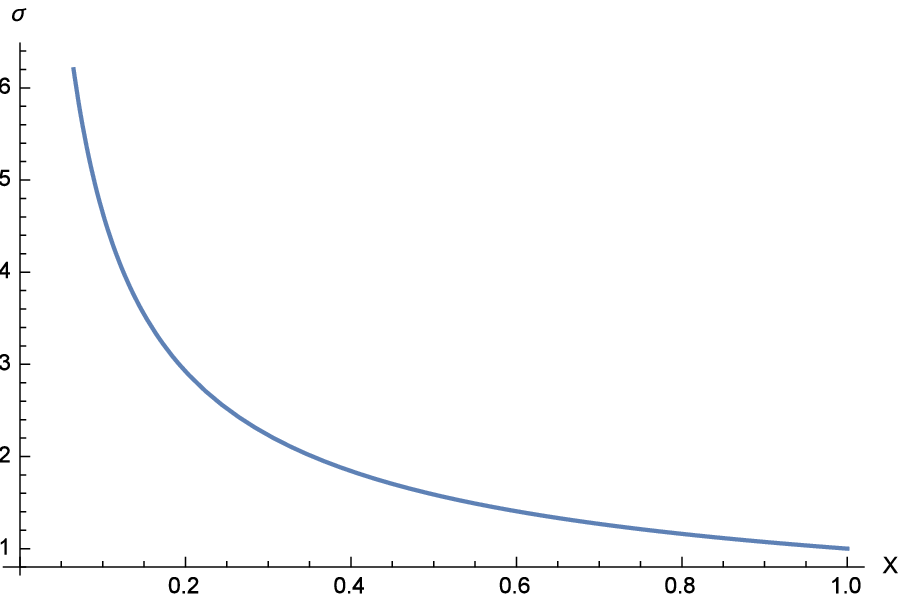}
        \caption{}
\end{subfigure}
\caption{The diagram showing (a) variation of Energy density $\rho$ with $\chi$,(b) variation of radiation energy $\epsilon$ with $\chi$, (c) variation of heat-flux $q$ with $\chi$, (d) variation of shear $\sigma$ with $\chi$, ($0< \chi <1$) } 
\label{fig1}
\end{figure}

\begin{figure}[h]
\begin{subfigure}[b]{0.6\textwidth}
\includegraphics[width=\textwidth]{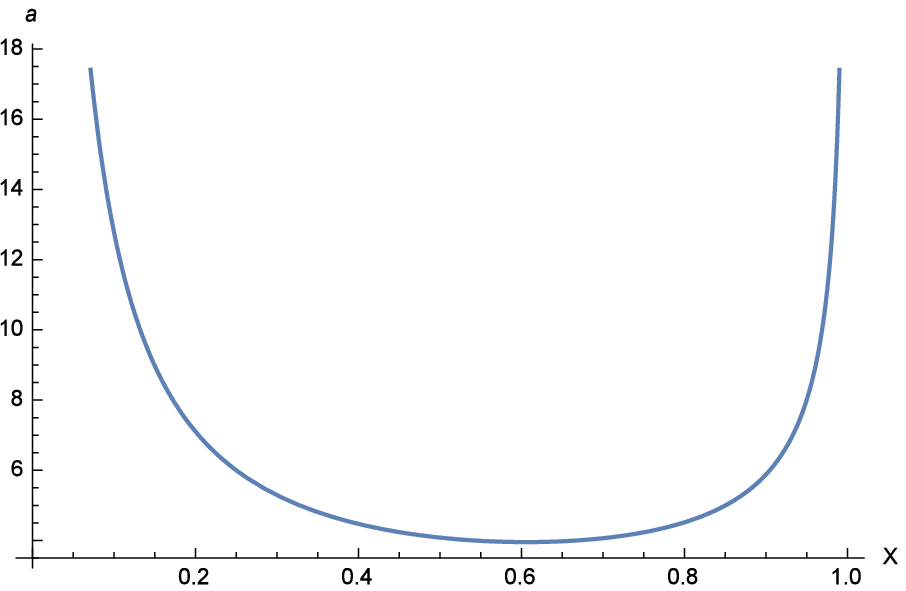}
        \caption{}
\end{subfigure}
\begin{subfigure}[b]{0.6\textwidth}
\includegraphics[width=\textwidth]{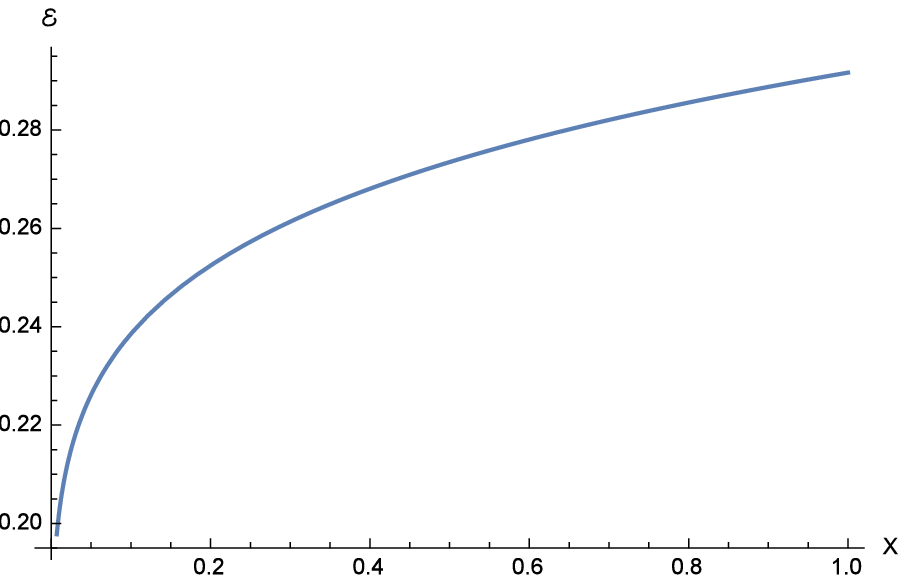}
        \caption{}
\end{subfigure}
\begin{subfigure}[b]{0.6\textwidth}
\includegraphics[width=\textwidth]{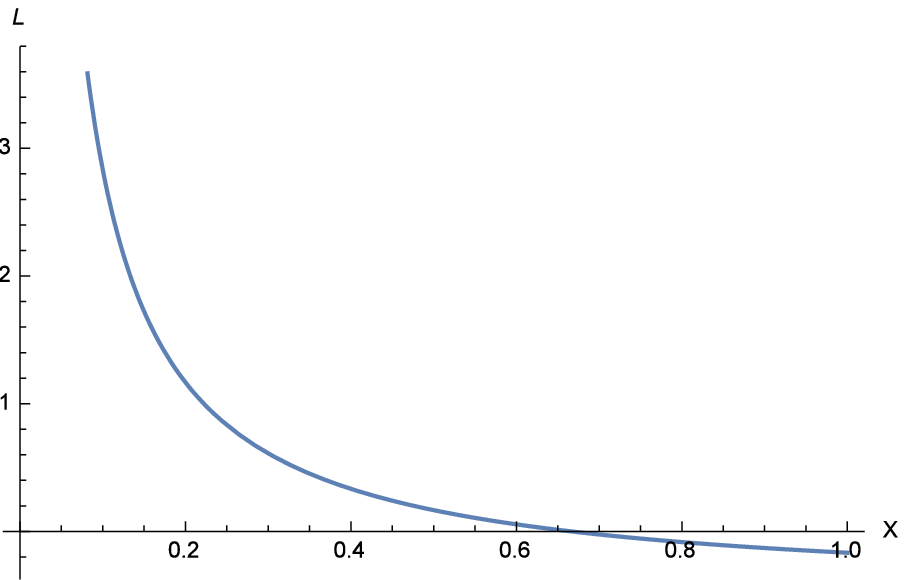}
        \caption{}
\end{subfigure}
\begin{subfigure}[b]{0.6\textwidth}
\includegraphics[width=\textwidth]{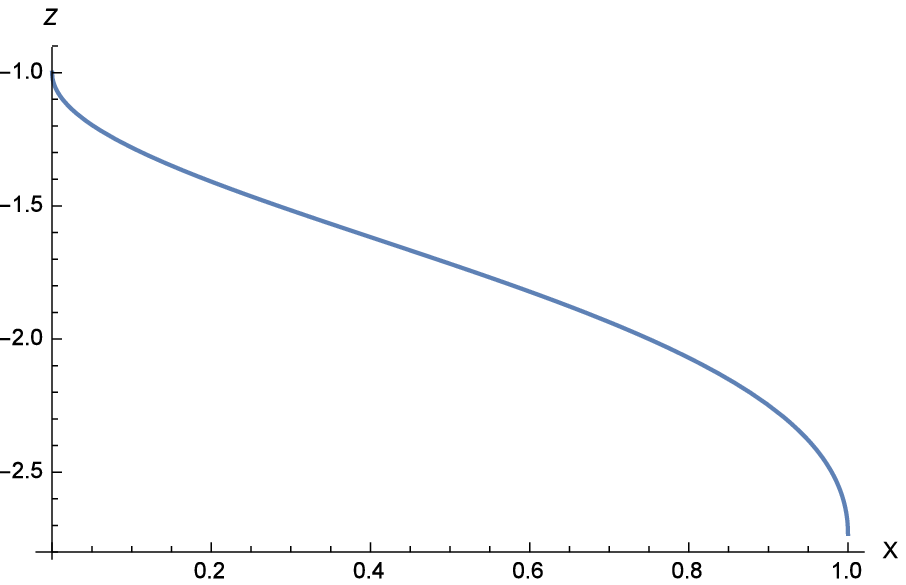}
        \caption{}
\end{subfigure}
\caption{The diagram showing (a) variation of acceleration $a$ with $\chi$, (b) variation of gravitational energy $\mathcal{E}$ with $\chi$, (c) variation of luminosity $L_{\Sigma^e}$ with $\chi$ and (d) variation of shift(blue-shift) $Z_{\Sigma^e}$ with $\chi$ observed at boundary surface. ($0< \chi <1$) } 
\label{fig2} 
\end{figure}

\section{Discussion and concluding remarks}
A new class of expansion free radiating was introduced by Herrera et al. \cite{lna08} which entails the existence of central cavity or, void. In recent, the evolution of such dynamical stars have taken considerable interest among relativists and has been applied to illustrate the physical and geometrical properties of radiating star. In our study, we have considered the self gravitating system of cylindrically symmetric dust dissipative fluids and examined that expansion free  radiating stars are rigorously constrained, and evolves with certain circumstances. It has explicitly shown that conformality of space time (Weyl Tensor) do not concerned by the presence of dissipation (heat flux and radiation energy) although Sherif et al.\cite{sgm19} showed that expansion free star are conformally flat under the condition of non-zero heat flux and  acceleration. Further, from the analysis of the system of Einstein's equation  we have proved that this class of dynamical star confined to the dissipation approximation (streaming out limit and diffusion approximation). It has also proved that a necessary  conditions to evolve such stars are that it must be radiating and accelerating\cite{sgm19}. 
\par
It is remarkable to note that very limited dynamical models are investigated for dissipative cavity evolution.  Therefore, we have also developed two families of solutions describing such evolution. Firstly, by considering the metric function as separation of variables to integrate analytically the relevant equations under different conditions, proving that such an integration could be achieved without undue difficulty. Although the obtained solutions were not very intended to illustrate the  astrophysical scenario but, it just bring out the potential of expansion free condition to model situations where central vacuum cavities are expected. Secondly, as an important geometrical case, self-similarity is introduced to discuss the cavity model. It is interesting to note here that solutions are explicitly described in terms of similarity variable $\chi$. This model has a space time singularity at $\chi = 0$ and the real analytical solution exists for $0< \chi <1$. The spectral radiation observed at boundary surface $\Sigma^e$ is blue-shifted ($Z_{\Sigma^e} <0$) and the luminosity of collapsing fluids as visible to observer vanishes for a fixed value $\chi = \frac{2}{3}$. The diagrams (\ref{fig1}) and (\ref{fig2}) showing the variation  of physical and kinematical parameters  with similarity variable $\chi$.


\end{document}